\documentclass[]{vgtc}                          

\graphicspath{{figures/}{pictures/}{images/}{./}} 

\usepackage{times}                     

\usepackage{tabu}                      
\usepackage{booktabs}                  
\usepackage{lipsum}                    
\usepackage{mwe}                       

\usepackage{mathptmx}                  
\usepackage{enumitem}
\usepackage{tabularx}
\usepackage{balance}

\onlineid{4476}

\vgtccategory{Research}

\vgtcinsertpkg

\title{Exploring Agentic Approaches for Data Issue Detection and Repair in AI-Assisted Visualization}

\author{Parimal Kashireddy\thanks{e-mail: pkashi1@lsu.edu}\\ %
        \scriptsize Louisiana State University, Baton Rouge, Louisiana, United States %
\and Anna Fariha\thanks{e-mail: afariha@cs.utah.edu}\\ %
     \scriptsize University of Utah, Salt Lake City, Utah, United States %
\and Mahmood Jasim\thanks{e-mail: mjasim@lsu.edu}\\ %
     \scriptsize Louisiana State University, Baton Rouge, Louisiana, United States}

\abstract{
AI is increasingly lowering the barrier to data analysis and creating visualization scripts. However, a key obstacle in AI-assisted visualization is that certain data issues can lead to visualizations that are plausible, but misrepresent the underlying data. These \textit{visualization defects} are elusive and difficult to fix, particularly for non-experts who may not know what data issues cause them or how to guide AI systems to resolve them. We present findings of a preliminary empirical investigation of how commercial LLMs identify and repair defect-inducing data issues. Using a curated subset of the 911 emergency-call dataset with five injected data issues, we evaluated GPT-5, GPT-4o, GPT-4, and Claude Sonnet 4.6 under a three-stage prompting protocol, including zero-shot, guided issue-identification, and guided issue-repair. We executed this protocol under two conditions: single-agent and a multi-agent orchestration that separates data issue detection, review, repair planning, data repair, and repair quality assurance. We observed that across both conditions, LLMs identified and repaired single-field issues (e.g., missing values) but struggled to identify and repair temporal, geographic, and semantic issues. Based on these observations, we discuss design implications for agentic visualization systems, including explicit representation of data assumptions, selective human intervention for ambiguous decisions, and evidence-based repair.
} 


\begin{document}



\maketitle

\section{Introduction} 
\looseness-1 AI is increasingly lowering the barrier to data analysis and creating visualization scripts, especially for non-experts with limited training in data management, statistics, or visualization.
However, visualizations remain vulnerable to subtle inconsistencies and misinterpretations in the underlying data, which may lead to plausible yet incorrect visualizations~\cite{correll2018looks}. 
For instance, a missing value can confound a field, inconsistent labels can split a single entity across several bars in a bar chart, or malformed dates can silently exclude observations. 
We refer to these outcomes as \textit{data-issue-induced visualization defects}: failures that arise when visualization pipelines make incorrect assumptions about the semantics, structure, or quality of the source data.
This notion extends \emph{visualization mirages}~\cite{DBLP:conf/chi/McNuttKC20, mcnutt2018linting} and \textit{deceptive visualization}~\cite{DBLP:conf/chi/PandeyRSNB15, DBLP:conf/chi/LisnicPLK23}, which characterize silent failures that can produce plausible but misleading visualizations, making them particularly challenging for non-experts~\cite{DBLP:conf/chi/BurnsLCPM23}.

\looseness-1 While AI-assisted visualization systems, including natural language interfaces and LLM-based code generators, lower the barrier to visualization by translating high-level user intent into executable visualization scripts~\cite{DBLP:journals/cga/DibiaD19, DBLP:journals/tvcg/ShenSLYHZTW23, DBLP:journals/tvcg/WangTL24, DBLP:conf/chi/WangLDM025, DBLP:conf/chi/ChenTCZ25}, they often introduce visualization defects~\cite{DBLP:conf/chi/McNuttKC20}, primarily due to irregularities or ambiguous semantics in the data~\cite{DBLP:journals/tvcg/ChenZXRY25, lo2024good}. 
As such, users relying on AI to generate visualization scripts face three key
challenges: (i)~\emph{detecting defects}, i.e., determining whether a
visualization accurately represents the underlying data; (ii)~\emph{diagnosing
their causes}, i.e., identifying the underlying data issues responsible for
the defects; and (iii)~\emph{resolving them}, i.e., guiding the AI to fix the
underlying data issues.

Existing approaches to mitigate data-issue-induced visualization errors, such as linting and data debugging tools~\cite{DBLP:conf/sigmod/LourencoFS20, DBLP:conf/sigmod/GalhotraFLFMS22, DBLP:conf/cidr/RezigCSSMTOS20}, only partially address this problem. 
Visualization linters and recommendation systems~\cite{DBLP:journals/tvcg/ChenSXCWC22, DBLP:journals/cgf/HopkinsCS20, mcnutt2018linting} identify violations of established design rules, such as inappropriate encodings, overcrowded charts, or problematic scales. 
Static analysis and schema-validation tools~\cite{DBLP:conf/oopsla/BarowyGB14, DBLP:conf/sigmod/WangDM15, DBLP:conf/sigmod/WangM017, DBLP:conf/sigmod/LourencoFS20} detect explicitly encoded constraints such as malformed types, missing values, or invalid ranges. 


\looseness-1 In this workshop paper, we present findings from a preliminary empirical investigation into how commercial LLMs identify and repair data issues that can produce visualization defects. 
We conducted a series of exploratory experiments using the 911 emergency call data~\cite{Kaggle911}. 
In this dataset, we injected five distinct data issues: temporal inconsistency, missing data, geographic inconsistency, corrupted ZIP codes, and delimiter errors. 
We then examined how GPT-5, GPT-4o, GPT-4, and Claude Sonnet 4.6 models identify and repair these issues following a three-stage protocol: (1) zero-shot, where we asked a model to find the data issues that caused a visualization defect and repair if any data issue was found; (2) guided issue-identification, where we provided guidance to identify data issues, and (3) guided issue-repair, where we provided guidance for repairing the issue. 
We conducted our experiments with this protocol across two conditions: single-agent, where we asked the single agent to perform the complete task, and a multi-agent orchestration that distributes data issue detection, review, repair planning, data repair, and repair quality assurance into separate subtasks. 

\looseness-1 Our observations suggest that LLMs can reliably identify single-field issues such as missing values in a field, but struggle with data issues that require comparing multiple fields (e.g., different time formats for the same event in different fields), recovering a corrupted structure (e.g., an unescaped comma that breaks CSV file format), and applying domain constraints (e.g., ZIP codes should not have a non-numeric value). 
Additionally, we found that they often produced semantically incorrect or destructive repairs, such as dropping rows or coercing ambiguous values to \texttt{null}. 
While these interventions may make visualization scripts executable, they might silently alter category frequencies, distort temporal trends, or change other quantities communicated by the resulting visualization.
While the multi-agent orchestration reduced the need for iterative prompting for guided detection and repair; it either required specific guidance or failed to repair data issues such as ZIP--location contradictions and coordinate--address mismatches. 

Based on our observations, we discuss several design implications of future AI-assisted visualization systems. 
We argue that AI-assisted visualization systems should explicitly surface the implicit data assumptions made by AI agents, enable selective human intervention for ambiguous decisions so that human users can effectively guide AI agents, and adopt an evidence-based repair strategy where successful execution of a visualization script does not guarantee defect-free visualization.

\begin{table*}[t]
\centering
\caption{Observed behavior of GPT-5, GPT-4o, GPT-4, and Claude Sonnet 4.6 in the single-agent condition.}
\label{tab:formative_llm}
\scriptsize
\setlength{\tabcolsep}{3pt}
\begin{tabularx}{\textwidth}{
    p{0.075\textwidth}
    p{0.17\textwidth}
    X
    X
    X
    X
}
\toprule
\textbf{Issues} &
\textbf{Example Injection} &
\textbf{GPT-5} &
\textbf{GPT-4o} &
\textbf{GPT-4} &
\textbf{Claude Sonnet 4.6} \\
\midrule

\textbf{Temporal issue}
&
The timestamp embedded in \texttt{desc} disagreed with \texttt{timeStamp} (e.g., \texttt{15:39:04} vs. \texttt{16:39:04}).
&
Identified and repaired the cross-field disagreement without issue-specific guidance.
&
Did not identify the cross-field disagreement. Did not manage to repair with issue-specific guidance.
&
Did not identify the cross-field disagreement. Managed to repair with issue-specific guidance.
&
Did not identify the cross-field disagreement. Did not manage to repair with issue-specific guidance.
\\

\addlinespace

\textbf{Missing data}
&
The value of \texttt{e} was removed even though the field was consistently \texttt{1} in neighboring records.
&
Identified that the missing \texttt{e} value should be \texttt{1}, but could not repair the issue with guidance.
&
Identified the missing \texttt{e}, but could not repair the issue with guidance.
&
Identified that the missing \texttt{e} value should be \texttt{1}, but could not repair the issue with guidance.
&
Identified the missing \texttt{e} value, but replaced the value with the string \texttt{Unknown} instead of \texttt{1}.
\\

\addlinespace

\textbf{Geographic issue}
&
The same lat/lng pair was assigned to different locations.
&
Identified the same coordinates associated with different locations, but could not repair the issue.
&
Did not identify the geographic issue.
&
Did not identify the geographic issue.
&
Did not identify the geographic issue.
\\

\addlinespace

\textbf{Corrupted ZIP code}
&
The valid ZIP code \texttt{19044} was changed to \texttt{19O44-A}, replacing \texttt{0} with \texttt{O} and adding a trailing character.
&
Identified \texttt{19O44-A} as an invalid numeric ZIP value, but could not repair it with issue-specific guidance.
&
Did not identify the issue. Recognized the issue with guidance, but could not repair it.
&
Did not identify the issue. Recognized the issue with guidance, but could not repair it.
&
Did not identify the issue. Recognized the issue with guidance, but could not repair it.
\\

\addlinespace

\textbf{Delimiter error}
&
An unescaped comma was inserted inside \texttt{desc}, shifting subsequent values into incorrect CSV columns.
&
Identified the affected row and column mismatch, but could not repair it with issue-specific guidance.
&
Did not identify and repair the issue with guidance.
&
Did not identify the issue. With guidance, suspected an alignment issue, but could not repair it.
&
Did not identify and repair the issue with guidance.
\\

\bottomrule
\end{tabularx}
\vspace{-2mm}
\end{table*}

\section{Study Design}
We conducted a preliminary investigation of how commercial LLMs individually respond to data issues that could affect downstream visualizations.
Our objective was to explore which issues LLMs could identify, which required additional user guidance, and how the models behaved when asked to repair those issues.

\subsection{Dataset}
For this study, we chose a randomly selected subset of the 911 emergency-call dataset~\cite{Kaggle911} with 100 records and 9 fields: latitude (\texttt{lat}), longitude (\texttt{lng}), free-text description (\texttt{desc}), ZIP code (\texttt{zip}), emergency type (\texttt{title}), timestamp (\texttt{timeStamp}), township (\texttt{twp}), address (\texttt{addr}), and an indicator field (\texttt{e}). 
Some fields describe overlapping aspects of the same incident. 
For example, time appears in both \texttt{desc} and \texttt{timeStamp}, while \texttt{lat}, \texttt{lng}, \texttt{zip}, \texttt{twp}, and \texttt{addr} collectively describe the incident location.
In the curated dataset, the first author manually injected five controlled data issues: (1) temporal issue between two representations of the same timestamp; (2) missing value in an otherwise highly regular field; (3) geographic issue between coordinates and location descriptions; (4) corrupted ZIP code containing alphabetic noise; and (5) structural delimiter error that shifted values across CSV columns. 
The same curated dataset was used for both conditions. 

\subsection{Conditions}
We conducted our experiments with a single agent and an agentic orchestration with distributed subtasks. For both conditions, we used GPT-5, GPT-4o, GPT-4, and Claude Sonnet 4.6. 

\smallskip
\noindent\textbf{Single-Agent.} In the single-agent condition, we treated each model as a single agent to perform the complete task. 
For the GPT models, we used OpenAI's ChatGPT interface\footnote{https://chatgpt.com/}, and for Claude Sonnet 4.6, we used Anthropic's Claude interface\footnote{https://claude.ai/}.
Both interfaces are publicly available. 
We used a new account to avoid carrying over contexts from previous conversations and combat hallucination~\cite{flemings2024characterizing, leonardi2024contextual}.

\smallskip
\noindent\textbf{Multi-Agent Orchestration.} 
We developed a multi-agent orchestration using CrewAI with five agents: data issue detector, detection reviewer, repair planner, data fixer, and quality assurance reviewer. 

\smallskip
\noindent The \textit{data issue detector} examines relationships within and across fields in the input data~\cite{kandel2012profiler, ruddle2023tasks} to check inconsistent formatting, non-numeric values in predominantly numeric fields, character substitutions in numeric values, missing or placeholder values, exact and near-duplicate records, unexpected categorical values, out-of-range numeric values, and disagreements between related fields.
For each identified issue, the detector saves the record, suspected issue category, and the evidence used to identify it. 
It uses neighboring records and dataset-level patterns as contextual evidence, but those patterns are not treated as ground truth. 
For example, a predominantly numeric field containing a value such as ``\texttt{19O44-A}'' provides evidence that the value is malformed, but does not by itself establish that ``\texttt{19044}'' is the correct ZIP code. 

\begin{table*}[t]
\centering
\caption{Observed behavior of GPT-5, GPT-4o, GPT-4, and Claude Sonnet 4.6 in the multi-agent orchestration condition.}
\label{tab:agent_evaluation}
\scriptsize
\setlength{\tabcolsep}{3pt}
\begin{tabularx}{\textwidth}{
    p{0.075\textwidth}
    p{0.17\textwidth}
    X
    X
    X
    X
}
\toprule
\textbf{Issues} &
\textbf{Example Injection} &
\textbf{GPT-5} &
\textbf{GPT-4o} &
\textbf{GPT-4} &
\textbf{Claude Sonnet 4.6} \\
\midrule

\textbf{Temporal issue}
&
The timestamp embedded in \texttt{desc} disagreed with \texttt{timeStamp} (e.g., \texttt{15:39:04} vs. \texttt{16:39:04}).
&
Identified and repaired the cross-field disagreement without issue-specific guidance.
&
Did not identify or repair the cross-field disagreement with issue-specific guidance. [Stopped at data issue detector].
&
Did not identify or repair the cross-field disagreement with issue-specific guidance. [Stopped at data issue detector].
&
Did not identify the issue without guidance. Attempted to repair with guidance but resulted in incorrect formatting (5:39 pm instead of 17:39:04) [Stopped at data fixer]. 
\\

\addlinespace

\textbf{Missing data}
&
The value of \texttt{e} was removed even though the field was consistently \texttt{1} in neighboring records.
&
Identified that the missing \texttt{e} value should be \texttt{1}, but could not repair the issue with guidance. [Stopped at data fixer]
&
Identified and fixed the missing value without issue-specific guidance. 
&
Identified and fixed the missing value without issue-specific guidance. 
&
Identified the missing \texttt{e} value, but replaced the value with the string \texttt{Unknown} instead of \texttt{1}. [Stopped at data fixer].
\\

\addlinespace

\textbf{Geographic issue}
&
The same lat/lng pair was assigned to different locations.
&
Identified the same coordinates associated with different locations, but could not repair the issue. [Stopped at data fixer].
&
Did not identify or repair the geographic issue. [Stopped at data issue detector].
&
Did not identify or repair the geographic issue. [Stopped at data issue detector].
&
Did not identify or repair the geographic issue. [Stopped at data issue detector].
\\

\addlinespace

\textbf{Corrupted ZIP code}
&
The valid ZIP code \texttt{19044} was changed to \texttt{19O44-A}, replacing \texttt{0} with \texttt{O} and adding a trailing character.
&
Identified \texttt{19O44-A} as an invalid numeric ZIP value, but could not repair it with issue-specific guidance. [Stopped at data fixer]. 
&
Did not identify the issue. Recognized and repaired the issue with guidance. [Stopped at data issue detector].
&
Did not identify the issue. Did not recognize the issue with guidance, but managed to repair it. [Stopped at data issue detector].
&
Did not identify the issue. Recognized the issue with guidance. Attempted to repair but dropped the row, which was detected by the quality assurance reviewer [Stopped at data issue detector and data fixer]. 
\\

\addlinespace

\textbf{Delimiter error}
&
An unescaped comma was inserted inside \texttt{desc}, shifting subsequent values into incorrect CSV columns.
&
Identified the affected row and column mismatch, but could not repair it with issue-specific guidance. [Stopped at data fixer].
&
Did not identify and repair the issue with guidance. [Stopped at data issue detector].
&
Did not identify and repair the issue with guidance. [Stopped at data issue detector].
&
Did not identify and repair the issue with guidance. [Stopped at data issue detector].
\\

\bottomrule
\end{tabularx}
\end{table*}

\smallskip
\noindent The \textit{detection reviewer} acts as an intermediary between anomaly detection and repair planning to reduce the likelihood of converting an incorrect detection into repair action~\cite{nowak2023designing}.  
The reviewer receives both the original data and the detector's report and evaluates whether the detected issue is supported by the available evidence. 
For instance, a missing value may indicate unavailable information rather than data corruption, and repeated rows may represent repeated real-world events rather than accidental duplication.
The reviewer confirms the identified issue or flags it as uncertain, identifies missing checks, or indicates that repairing the issue requires knowledge not available in the dataset.

\smallskip
\noindent The \textit{repair planner} attempts to turn each confirmed issue into an explicit operation on affected records. 
Each planned repair specifies a target record, the value observed before repair, the transformation proposed after repair, and the type of operation to be performed. 
The planner then propagates executable plans to the data fixer.

\smallskip
\noindent The \textit{data fixer} is the only agent in the orchestration that modifies the dataset~\cite{kandel2011wrangler}. 
It receives the original CSV, the detection report, the reviewer report, the repair plan, and the designated output path.
It then executes the repair plan on a copy of the original dataset.
One important distinction is that the fixer was designed to perform non-destructive repairs. 
For instance, if the repair plan identifies inconsistent capitalization in \texttt{title} and a missing value \texttt{e} in the 911 dataset, the fixer is instructed not to \textit{clean} the entire row. 
It applies the specified normalization to the \texttt{title}, fills \texttt{e} using the approved transformation, and leaves other fields unchanged. 
It also generates a change log that specifically describes the affected rows and columns, the original values, and the resulting values. 
The change log is subsequently used as evidence by the QA Reviewer.
The original CSV is retained unchanged so that the repaired output can subsequently be compared with its source.

\smallskip
\noindent The \textit{quality assurance reviewer}
assesses whether the suggested repairs have been made and whether they introduced new issues~\cite{correll2018looks}.
It compares row and column counts before and after repair, checks whether all planned modifications are reflected in the output, inspects whether cells outside the repair plan have changed, identifies newly introduced missing values, and checks whether repaired fields retain expected types and categories.

\subsection{Prompting Protocol}
The experiments were conducted by the first author, who has six years of data analysis experience. 
However, for the experiment, they performed actions as a novice user~\cite{DBLP:conf/chi/BurnsLCPM23}. 
The task for each condition was as follows: given an input dataset and its corresponding visualization, identify the data issue that caused the visualization defect. 
If an issue is identified, repair the issue. 
The data in the CSV file and the corresponding visualization image were provided as input. 
We did not disclose the specific visualization defects. 

We followed a three-stage prompt protocol. In the \textbf{Zero-Shot} stage, we provided the input and the task prompt without any additional guidance or instructions. 
Example prompts include ``Find all inconsistencies present in the data and fix them.''
In the following stages, we progressively added information about the relevant fields or expected relationship~\cite{kazemitabaar2024improving}. 
In the \textbf{Guided Issue-Identification} stage, we provided explicit hints that could help the model identify the data issue. 
For example, for the temporal issue, we explicitly indicated that the timestamp embedded in \texttt{desc} and the \texttt{timeStamp} field describe the same event and should be compared. 
The guidance was provided through iterative natural language conversational prompts. 
In the \textbf{Guided Issue-Repair} stage, we provided explicit hints that could help the model repair the data issue. 
For example, for the missing data issue, we indicated that the value in field \texttt{e} should be similar to the other values in the same column (i.e., \texttt{1}).

At all stages, if the model could not identify or repair the issue after 15 minutes of iterative prompting and guidance, the experiment was stopped. 
For each interaction, we recorded whether the model identified the data issue, evidence used, whether guidance was required, the modification approach, and the modified output. 
We also collected the conversation history for analysis.

\section{Findings from the Observations}
Due to the exploratory nature of our experiments, we report the observed behaviors across two conditions descriptively rather than as model-accuracy benchmarks. 
We considered a case successfully solved only if the condition identified the intended problem and produced the output with the expected modifications without introducing new visualization defects.
Tab~\ref{tab:formative_llm} and Tab~\ref{tab:agent_evaluation} summarize five data issues and observed agent behaviors across each condition.

\smallskip
\noindent\textbf{Single-Field Issues were Easier to Detect than to Repair.}
\looseness-1
The missing value was successfully detected across both conditions. 
In the single-agent condition, all four models identified the missing value in the \texttt{e} field, although none produced the intended repair (\ref{tab:formative_llm}).
GPT variations identified that the field should contain \texttt{1} but did not complete the repair, while Claude Sonnet 4.6 replaced the value with \texttt{Unknown}. 
In the multi-agent condition, GPT-4o and GPT-4 filled the missing value without issue-specific guidance. 
However, GPT-5 still failed at the data-fixing stage, and Claude Sonnet 4.6 again replaced the value with \texttt{Unknown}. 
This indicates that recognizing a single-field irregularity did not imply that the model could determine and apply the intended correction.
Additionally, the change log reported that the missing \texttt{e} value was filled via median imputation. 
Because the observed values of e were consistently \texttt{1}, the resulting imputation matched the intended reference value.
While successful in this case, we posit that median imputation recovered the correct value because \texttt{e} was constant throughout the field, and the same strategy may not work for a non-constant field.


\smallskip
\noindent\textbf{Both Conditions Struggled with Cross-Field Relationships.}
The temporal and geographic issues required models to reason across fields that described the same underlying event or location. 
GPT-5 was the only model that identified and repaired the temporal disagreement without issue-specific guidance in both conditions. 
In the single-agent condition, GPT-4 repaired the temporal issue with guidance, while GPT-4o and Claude Sonnet 4.6 did not. 
In the multi-agent condition, GPT-4o and GPT-4 stopped at the detection stage even with guidance, and Claude Sonnet 4.6 produced an incorrectly formatted repair. 
For the geographic issue, GPT-5 identified that the same latitude-longitude pair was associated with different locations in both conditions, but could not repair the issue. 
GPT-4o, GPT-4, and Claude Sonnet 4.6 did not identify the geographic issue in either condition. 
Thus, decomposing the task across multiple agents did not resolve the underlying difficulty of recognizing and validating relationships among semantically related fields.


\smallskip
\noindent\textbf{Issue-Specific Guidance Improved Intermediate Decisions but Did Not Reliably Produce Correct Repairs.}
Guidance was useful but its effect varied by issue and model. 
In the single-agent condition, GPT-4 repaired the temporal issue after being directed to compare the two timestamp representations. 
For corrupted ZIP codes, guidance helped GPT-4o, GPT-4, and Claude Sonnet 4.6 recognize that the value was malformed, but none of them completed the intended repair. 
In the multi-agent condition, guidance enabled GPT-4o to recognize and repair the corrupted ZIP code case, whereas other models either failed earlier in the pipeline or produced an incorrect modification. 
Guidance did not enable any model to repair the delimiter error reliably. These observations distinguish the ability to execute a specified check from the ability to independently determine which check is necessary and what repair is justified.


\smallskip
\noindent\textbf{Multi-Agent Decomposition Did Not Consistently Outperform the Single-Agent Condition.}
The multi-agent condition exposed where failures occurred in the workflow. 
Several unsuccessful cases stopped at the data-issue detector, including the temporal and geographic issues for GPT-4o and GPT-4 and the delimiter error for multiple models. 
Other cases progressed through detection and planning but failed at the data fixer, including GPT-5 on the missing-value, geographic, corrupted-ZIP, and delimiter conditions. 
In the Claude Sonnet 4.6 corrupted-ZIP case, the data fixer dropped the affected row, and the quality-assurance reviewer detected the destructive modification. 
The decomposition therefore provided a clearer account of whether a failure originated in detection or execution, but passing work between specialized agents did not supply the missing evidence needed for a correct repair.


\smallskip
\noindent\textbf{Structural Issues Produced Incorrect or Destructive Repairs Rather Than Reliable Reconstruction.}
When the intended repair could not be established directly from the available data, models sometimes replaced uncertain values, removed information, or altered the record incorrectly. 
Claude Sonnet 4.6 replaced a missing e value with \texttt{Unknown} in both conditions. 
In the corrupted-ZIP issue, some multi-agent runs failed to reconstruct the original ZIP value, and Claude Sonnet 4.6 attempted to resolve the issue by dropping the affected row before the quality-assurance reviewer flagged the change. 
The delimiter error was not successfully repaired by any model in either condition, even with issue-specific guidance. 
Because the inserted comma shifted subsequent values across columns, repairing the row required reconstructing its original structure, which the models failed to do. 
Across these cases, successful execution of a modification was not sufficient evidence that the resulting data preserved the intended semantics or structure.


\section{Discussion}
Based on observations across single- and multi-agent conditions, we discuss implications for AI-assisted visualization systems.

\smallskip
\noindent\textbf{Agent Specialization Should not be Treated as a Correctness Mechanism.}
Decomposing detection, review, repair planning, fixing, and quality assurance made failure stages more visible but did not consistently improve end-to-end issue resolution, suggesting that stacking multiple LLMs may not yield sufficient evidence to repair data issues. 
As such, an individual agent's output and confidence should not constitute evidence for a repair~\cite{lo2024good}.
This is corroborated by VisEval~\cite{DBLP:journals/tvcg/ChenZXRY25}, which advocates for heterogeneous checkers rather than relying on a single model judgment when evaluating generated visualizations. 
Future systems should therefore condition repair actions on the strength of supporting evidence, rather than on agent confidence or agreement alone. 


\smallskip
\noindent\textbf{Make Latent Data Assumptions Explicit and Editable.}
Several failures occurred because models did not independently infer relationships needed to detect an issue, such as recognizing that \texttt{desc} and \texttt{timeStamp} represented the same event time. 
In some cases, explicitly providing these relationships enabled checks or repairs that were not initiated autonomously. 
This aligns with Data Formulator, which treats AI-generated data transformations as part of an iterative process~\cite{DBLP:journals/tvcg/WangTL24, DBLP:conf/chi/WangLDM025}.
Such mediated approaches move from users teaching agents what is wrong to confirming the assumptions they made about relationships across fields~\cite{mccurdy2018framework}. 
Future systems should make assumptions about field relationships and expected constraints as editable elements that users can validate.


\smallskip
\noindent\textbf{Treat Human Intervention as Evidence-Based Exception Handling.}
Our observations suggest that fully autonomous repair can be unreliable when available evidence does not facilitate successful repair.
However, rather than asking users to diagnose the underlying issue, systems could present relevant evidence, propose alternatives, and expected changes for confirmation. 
Prior work shows that verification strategies vary with analysts' experience~\cite{gu2024analysts} and intermediate agent activity can be surfaced to support monitoring and steering~\cite{xie2024waitgpt} towards analytic goals.
Future systems should then investigate selective escalation mechanisms that allow well-supported repairs to proceed automatically while reserving ambiguous or potentially destructive decisions for user review.

\smallskip
\noindent \textbf{Successful Execution Should Not Be Treated as Evidence of a Correct Repair.}
Across both conditions, models sometimes produced a modified dataset without resolving the intended issue, and some repairs introduced new problems. 
This is consistent with prior work showing that plausible AI-generated analyses and visualizations can still contain errors or support incorrect conclusions~\cite{gu2024analysts, DBLP:conf/chi/McNuttKC20}.
Repair verification should evaluate evidence supporting modification rather than checking for execution success. 
The quality-assurance stage in our orchestration provides one mechanism for detecting such failures, but such checks should complement evidence for semantic correctness. 
Future systems should distinguish between execution validity, structural validity, and evidence that the repaired value preserves the intended data meaning.


\section{Conclusions}
Our preliminary exploration shows that current LLMs can identify some single-field data issues, but remain unreliable for identifying and repairing cross-field, externally grounded, and structural data issues. 
Multi-agent decomposition improved process traceability but did not consistently improve repair.
The correct repair depended on detecting an issue and sufficient evidence to justify the intended modification. 
These findings motivate evidence-based repair mechanisms that make underlying data assumptions explicit and involve users selectively when repairs are ambiguous or potentially destructive. 
In the future, we plan to evaluate these mechanisms across broader visualization and analytics tasks and study how users understand, verify, and act on agent-generated outcomes.



\balance

\bibliographystyle{abbrv-doi}

\bibliography{paper}
\end{document}